\documentclass{ifacconf}

\usepackage{graphicx}      
\usepackage{natbib}        

\usepackage{cmap}
\usepackage{xcolor}
\usepackage{amsfonts}
\usepackage{amsmath}
\usepackage{amssymb}

\newtheorem{theorem}{Theorem}

\begin{document}
\begin{frontmatter}

\title{Divergence Method to Stability Study of  Andronov-Vyshnegradsky Problem. Hidden Oscillations}


\author[First]{I.B. Furtat} 
\author{N.V. Kuznetsov} 

\address[First]{IPME RAS, St. Petersburg, Boljshoy pr. V.O., 61, Russia (e-mail: cainenash@mail.ru).}

\begin{abstract}                
The classical Andronov-Vyshnegradsky problem, which deals with locating regions of stability and oscillations in control systems with a Watt regulator, is solved using a divergence method for studying the stability of dynamic systems.
This system is studied both with and without the self-regulation effect.
The exact value of the hidden boundary of the global stability region is obtained.
The stability criteria for a system with a Watt regulator are also presented in the context of the solvability of a linear matrix inequality.
Computer modelling shows that the system exhibits hidden oscillations when the self-regulation effect is present and when it is not.
The conditions for computing the hidden boundary of global stability are determined by three parameters in the Watt regulator model.
\end{abstract}


\end{frontmatter}

\section{Introduction}

Watt's centrifugal governor (or Watt regulator) is one of the key inventions of the Industrial Revolution.
It is a mechanical device that automatically regulates the speed of a steam engine, making it safe and efficient.
Detailed information on the operation of Watt's governor can be found, for example, in \cite{Vyshnegradsky1876,farcot1873,mikerov2014}.
Watt's governor acts as a simple mechanical feedback loop, maintaining a constant engine speed without direct human intervention.
Modifications of Watt's governor in \cite{farcot1873} can still be found in alternators with automatic voltage regulation, in wind turbines (hurricane protection systems), and in marine diesel engines (Woodward governor) \cite{denny2002}.
Until now, the operation of Watt's regulator raises a lot of problems and ways of studying them, see, for example \cite{albu2021,schaper2023,baharin2024,silva2022,kowalczyk2025}.

Research into the operation of Watt's regulator led to the Andronov-Vyshnegradsky problem, see \cite{andronov1945}, which is a cornerstone of the theory of nonlinear oscillations and automatic control.
After the introduction of Watt's regulator, engineers encountered a problem that was, at the time, inexplicable: on some steam engines, the regulator worked perfectly, maintaining a constant speed. On others, it began to "oscillate": the engine speed would either increase or decrease excessively, causing strong oscillations that could lead to machine failure.
These unwanted oscillations are self-excited oscillations, or, in some cases, undamped oscillations, see \cite{maxwell1868}.

The Andronov-Vyshnegradsky problem is to determine the system parameters at which its operation is stable (oscillations decay, the speed reaches a constant value).
It is required to determine the stability boundary, where the system operates near the threshold where oscillations break down. 
Also, it is needed to predict the parameters at which stable self-oscillations occur (limit cycle).

The analysis of stability and the occurrence of oscillations in control systems is one of the central problems of mathematical control theory, where the Andronov-Vyshnegradsky problem is one of the main problems for study, see \cite{Andrievsky2021,Pokrovskiy2022,Chen2023,Smirnova2023,Wang2024}.
In recent years, the development of this area within the framework of hidden oscillation theory has been associated with finding the exact boundary of global stability in the parameter space of a nonlinear control system, see \cite{kuznetsov2020}.
To obtain an internal (conservative) estimate of the boundary of global stability, Lyapunov's methods were developed (see, for example, \cite{lurie1944,barbashin1952,lasalle1961,gelig1978}).
To obtain an external estimate, various methods for analyzing the origin of periodic oscillations have been proposed (see, for example, \cite{vanderpol1920,bogoliubov1937,tsypkin1984,boiko2021}).

In many cases, the difficulties in solving such problems are associated with the possible existence of hidden attractors in the phase space of the system and hidden parts of the global stability boundary in the parameter space (see \cite{kuznetsov2020,kuznetsov2020ecc,kuznetsov2022,kuznetsov2023ifac}).

In \cite{kuznetsov2022,kuznetsov2023ifac}, a solution to the Andronov-Vyshnegradsky problem using the Lyapunov function method is considered. 
However, in \cite{kuznetsov2022,kuznetsov2023ifac} it is shown that the method of Lyapunov functions does not allow one to determine the boundaries and regions of stability.

This paper proposes a solution to the Andronov - Vyshnegradsky problem to find the boundaries and regions of stability using a novel divergence method for studying stability (see \cite{Furtat20a,Furtat20b,Furtat21}).
For the case of the absence of self-regulation in the Watt regulator, the divergence method confirms the results obtained in \cite{kuznetsov2022,kuznetsov2023ifac}.
Additionally, the divergence method enables the study of stability in the presence of self-regulation, yielding a system stability criterion expressed as the solvability of a linear matrix inequality (LMI).
It has been demonstrated that hidden oscillations exist in a system with a Watt regulator, both in the absence and presence of the self-regulation effect.

The paper is organized as follows. 
Subsection \ref{SubSec11} introduces the basic notations and definitions. 
Section \ref{Sec_Main_Result} presents the Watt regulator model and considers the Andronov-Vyshnegradsky problem. 
Section \ref{Sec_Main_Result} also briefly summarizes the necessary and sufficient stability conditions based on the divergence method. 
Section \ref{Density} considers the main results. 
The simulation, confirming the theoretical conclusions, are also presented in Section \ref{Density}. 
Section \ref{Concl} contains the main conclusions.


\subsection{Notations and Definitions}
\label{SubSec11}

The following \textit{notations and definitions are used in this paper:}
$\mathbb R^n$ is the $n$-dimensional Euclidean space,
$\mathbb R^{n \times n}$ is the set of all real matrices of dimension $n \times m$,
``$*$'' denotes the symmetric block of a symmetric matrix,
$P>0$ ($P<0$) means that $P$ is positive (negative) definite,
$P \geq 0$ ($P \leq 0$) means that $P$ is positive (negative) semidefinite,
$grad \{W(x)\}=\Big[ \frac{\partial W}{\partial x_1}, ..., \frac{\partial W}{\partial x_n} \Big]^{\rm T}$ is the gradient of a scalar function $W(x)$,
$div\{h(x)\}=\frac{\partial h_1}{\partial x_1}+...+\frac{\partial h_n}{\partial x_n}$ is the divergence of the vector field $h(x)=[h_1(x),...,h_n(x)]^{\rm T}$.

According to \cite{kuznetsov2020,kuznetsov2022,kuznetsov2023ifac}, in general, the global stability boundary is the closure of the set of points in the system's parameter space for which the system is not globally stable.
Points on the global stability boundary are bifurcation points where undamped oscillations arise.
A boundary point is called \textit{hidden} if, for its arbitrary neighborhood in the parameter space, the loss of global stability is caused only by global bifurcations of the generation of hidden oscillations, the region of attraction of which in the phase space does not touch any of the unstable equilibrium states. Otherwise, the point is called trivial.


\section{Watt's Regulator Model. Preliminary Information on the Divergent Stability Method}
\label{Sec_Main_Result}

\subsection{Watt's Governor Model}

Applying Newton's laws for rotational motion, taking into account dry friction and self-regulation, and following \cite{Vyshnegradsky1876}, assuming that the governor sleeve displacement is small, we write equations describing the governor dynamics as follows
\begin{equation}
\label{eq_Watt}
\begin{array}{lll}
\ddot{w}(t)+B\dot{w}(t)+Aw(t)=u(t)-k \textup{sign}(\dot{w}(t)),
\\
\dot{u}(t)+Cu(t)=-w(t).
\end{array}
\end{equation}
The first equation in \eqref{eq_Watt} describes the dynamics of the controller's motion, the second describes the dynamics of the drive (motor). The following notations are used:
$t \geq 0$ is a time,
$w \in \mathbb R$ is the main coordinate of the governor (usually this is the distance between the governor weights (balls) or the position of the coupling connected to them; $w$ is used to directly control the valve that supplies steam to the engine),
$u \in \mathbb R$ is the control action (e.g., the power or torque developed by the engine),
$A>0$ is the coefficient of "stiffness" or restoring force (this constant is determined by the governor design, for example, the stiffness of the springs, which tends to return the balls to some "equilibrium" position),
$B>0$ is the damping coefficient (viscous friction; this constant describes the resistance forces proportional to the speed, for example, friction in the hinges, air resistance; it dampens governor oscillations),
$C>0$ is the coefficient reciprocal of the drive time constant ($C=0$ corresponds to the case of the absence of self-regulation, and the system \eqref{eq_Watt} coincides with that considered in \cite{Vyshnegradsky1876}),
$k \textup{sign}(\dot{w}(t))$ is a dry friction model with a dry friction coefficient of $k$.
As in \cite{kuznetsov2022,kuznetsov2023ifac}, we set $k=0.5$.
The solutions of \eqref{eq_Watt} are understood in the sense of Filippov (\cite{filippov1985}), and numerical studies of such systems are carried out using the procedures from \cite{piiroinen2008}.

Rewrite the system \eqref{eq_Watt} in Lurie form (\cite{lurie1944}) as follows
\begin{equation}
\label{eq_Lurie}
\begin{array}{lll}
\dot{x}_1=x_2,
\\
\dot{x}_2=-Ax_1-Bx_2+x_3+0.5 \varphi(x_2),
\\
\dot{x}_3=-x_1-Cx_3,
\end{array}
\end{equation}
where $ \varphi(x_2)=\textup{sign}(x_2)$, or in matrix form
\begin{equation}
\label{eq_Lurie0}
\begin{array}{lll}
\dot{x}=Hx+G \varphi(\sigma),~~~\sigma=Lx,
\end{array}
\end{equation}
where $x = col\{x_1,x_2,x_3\}$ is the state, $\varphi(\sigma)=\textup{sign}(\sigma)$,
$H=
\begin{pmatrix}
0 & 1 & 0 \\
-A & -B & 1 \\
-1&0&-C
\end{pmatrix}$,
$G= (0~0.5~0)^{\rm T}$,
$L=(0~1~0)^{\rm T}$.

\subsection{Divergent Method}

Consider a dynamic system in the form
\begin{equation}
\label{eq1}
\begin{array}{l}
\dot{x}=f(x),
\end{array}
\end{equation}
where $x=[x_1, ..., x_n]^{\rm T}$ is the state, $f=[f_1,...,f_n]^{\rm T}: D \to \mathbb R^{n}$ is a continuously differentiable function defined in the domain $D \subset \mathbb R^{n}$. The set $D$ contains the origin and $f(0)=0$. We denote $\bar{D}$ as the boundary of the domain $D$.


Consider a necessary stability condition \cite{Furtat20a,Furtat20b,Furtat21}.
\begin{theorem}
\label{Th2}

Let $x=0$ be an asymptotically stable equilibrium point of the system \eqref{eq1}.
Then there exists a positive-definite continuously differentiable function $\rho(x)$ such that $\rho(x) \to \infty$ as $x \to \bar{D}$ and one of the following conditions holds:
\begin{enumerate}
\item[(i)] the function $div\{\rho(x)f(x)\}$ is integrable in the domain $V=\{x \in D: S(x) \leq C\} \subset D$ and $\int_{V} div\{\rho(x)f(x)\} dV < 0$ for all $C>0$;

\item[(ii)] the function $div\{\rho^{-1}(x)f(x)\}$ is integrable in the domain $V_{inv}=\{x \in D: S^{-1}(x) \geq C\} \subset D$ and $\int_{V_{inv}} div\{\rho^{-1}(x)f(x)\} dV_{inv} > 0$ for all $C>0$.
\end{enumerate}
\end{theorem}

Consider a sufficient stability condition \cite{Furtat20a,Furtat20b,Furtat21}.

\begin{theorem}
\label{Th2a}
Let $\rho(x)$ be a positive-definite continuously differentiable function defined in the domain $D$. Then the point $x=0$ is stable (asymptotically stable) if one of the following conditions is satisfied:
\begin{enumerate}
\item[(i)] $div\{\rho(x)f(x)\} \leq \rho(x) div\{f(x)\}$ ($div\{\rho(x)f(x)\} < \rho(x) div\{f(x)\}$) for any $x \in D \setminus \{0\}$;

\item[(ii)] $div\{\rho^{-1}(x)f(x)\} \geq 0$ ($div\{\rho^{-1}(x)f(x)\} > 0$) and $div\{f(x)\} \leq 0$ for any $x \in D \setminus \{0\}$;

\item[(iii)] $div\{\rho(x)f(x)\} \leq \beta(x) \rho^2(x) div\{\rho^{-1}(x)f(x)\}$ 

($div\{\rho(x)f(x)\} < \beta(x) \rho^2(x) div\{\rho^{-1}(x)f(x)\}$), where $\beta(x) > 1$ and $div\{f(x)\} \leq 0$ or only $\beta(x) = 1$ for any $x \in D \setminus \{0\}$.
\end{enumerate}
\end{theorem}


\section{Application of the Divergent Method for Stability Study}
\label{Density}

As in \cite{kuznetsov2023ifac}, we study the system \eqref{eq_Lurie} outside the discontinuity surface $S=\{(x_1,x_2,x_3) \in \mathbb R^3: x_2 = 0\}$.

\subsection{Obtaining an exact hidden boundary for global stability}
\label{Density1}

Choose the density function $\rho(x)$ as follows
\begin{equation}
\label{eq_Density}
\begin{array}{lll}
\rho(x)= & 0.5A\left[ \frac{1}{A}x_1+x_2 \right]^2
\\
&+0.5 \left[ Ax_1-x_3+\frac{AB-1}{A}x_2+0.5\varphi(x_2) \right]^2
\\
& +\frac{AB-1}{2A^2}x_2^2+0.5 \alpha x_3^2.
\end{array}
\end{equation}

Check that the necessary stability condition (see Theorem \ref{Th2}) is satisfied.
Taking into account \eqref{eq_Lurie} and \eqref{eq_Density}, we calculate
\begin{equation}
\label{eq_Density_Necc_1}
\begin{array}{lll}
div\{\rho(x)f(x)\}= & -\frac{AB-1}{A} 
\\
& \times \left[x_3-Ax_1-Bx_2-0.5\varphi(x_2) \right]^2
\\
& -\alpha x_1 x_3 - \alpha C x_3^2 - (B+C)\rho(x)
\\
& = z^{\rm T} Wz - (B+C)\rho(x),
\end{array}
\end{equation}
where
$W=W^{\rm T}=(W_{ij})$,
$W_{11}=-A(AB-1)$,
$W_{12}=-B(AB-1)$,
$W_{13}=AB-1-0.5\alpha+0.5AC$,
$W_{14}=0.5(AB-1)$, 
$W_{22}=-\frac{B^2(AB-1)}{A}$,
$W_{23}=-\frac{C}{2A}+0.5CB-\frac{B(AB-1)}{A}$,
$W_{24}=\frac{B(AB-1)}{2A}$,
$W_{33}=-C-\frac{AB-1}{A}-\alpha C$, 
$W_{34}=-0.25C-\frac{AB-1}{2A}$, 
$W_{44}=-\frac{AB-1}{4A}$, 
$z=col\{x,\varphi\}$.

The inequality $div\{\rho(x)f(x)\}<0$ is satisfied for $\alpha=0$ and $C=0$.
Therefore, the first condition (i) in Theorem \ref{Th2} is satisfied.

Let us now check the sufficient stability conditions (see Thorem \ref{Th2a}).
Taking into account \eqref{eq_Lurie} and \eqref{eq_Density}, calculate the following expressions
\begin{equation}
\label{eq_Density_Case1_1}
\begin{array}{lll}
div\{\rho(x)f(x)\}-\rho(x)div\{f(x)\}=
\\
- \frac{AB-1}{A} \left[x_3-Ax_1-Bx_2-0.5\varphi(x_2) \right]^2  
\\
-\alpha x_1 x_3 - \alpha C x_3^2 =
z^{\rm T} W z,
\\
\\
div\{\rho(x)f(x)\}-\rho^2(x)div\{\rho^{-1}(x)f(x)\} 
\\
= 0.5 z^{\rm T}Wz.
\end{array}
\end{equation}

If $AB>1$, $\alpha=0$, and $C=0$, then $div\{\rho(x)f(x)\}-\rho(x)div\{f(x)\}<0$ and $div\{\rho(x)f(x)\}-\rho^2(x)div\{\rho^{-1}(x)f(x)\}<0$.
Therefore, the conditions (i) and (iii) of Theorem \ref{Th2a} are satisfied.

Thus, the hidden boundary of global stability for $C=0$ is defined by the expression $AB=1$.
When $C > 0$, theoretical analysis of the matrix $W$ is difficult, so we will carry it out using computer modeling.


\subsection{Numerical Studies of Results of Section \ref{Density1}}

If $C=0$ and $\alpha=0$, then the calculation of the global stability region by checking the condition $W>0$ through simulation coincides with the theoretical result $AB>1$.

In turn, the boundary $AB=1$ is a hidden boundary of global stability.

If $C>0$ and $\alpha>0$, then simulation shows that regions of instability may appear in the region $AB>1$.

In the next section, we will show that these regions are actually absent, and they appear due to the conservatism of the solution obtained in Section \ref{Density1}.

\begin{figure}[h]
\center{\includegraphics[width=0.9\linewidth]{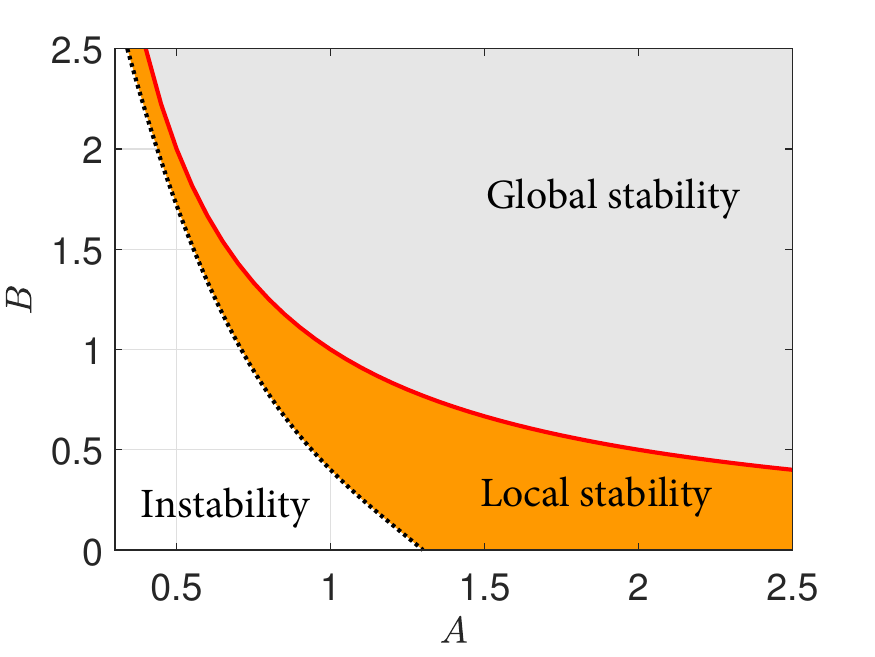}}
\caption{Stability regions with $C=0$ and $\alpha=0$ for parameters $A$ and $B$.}
\label{Fig_Exact}
\end{figure}

Fig.~\ref{Fig_Exact} shows the global stability region for $C=0$.
Regions of local stability and instability were found in \cite{boiko2021,kuznetsov2023ifac} using the Tsypkin method and the harmonic balance method.

\subsection{Global stability region as a solvability of a linear matrix inequality (LMI)}
\label{Density2}

Now we choose the density function $\rho(x)$ in the form
\begin{equation}
\label{eq_Density_new}
\begin{array}{lll}
\rho(x)=x^{\rm T}Qx,
\end{array}
\end{equation}
where $Q=Q^{\rm T}>0$ and $Q \in \mathbb R^{3 \times 3}$.

Let us check that the necessary stability condition (see Theorem \ref{Th2}) is met.
Taking into account \eqref{eq_Lurie} and \eqref{eq_Density}, we calculate
\begin{equation}
\label{eq_Density_new_Necc_1}
\begin{array}{lll}
div\{\rho(x)f(x)\} = & x^{\rm T}(H^{\rm T}Q+QH)x
\\
& +2x^{\rm T}BH\varphi(x_2)
\\
& =z^{\rm T} F_0 z 
- (B+C)\rho(x),
\end{array}
\end{equation}
where
$F_0=\begin{pmatrix}
H^{\rm T}Q+QH & BH \\
*&0
\end{pmatrix}$.
Since $\varphi^2(x_2) \leq 1$, then
$z^{\rm T}\begin{pmatrix}
0 & 0\\
*&-1
\end{pmatrix}
z\leq 1$.
Using the S-procedure, if the inequalities $div\{\rho(x)f(x)\}<0$ and $\varphi^2(x_2) \leq 1$ hold simultaneously, then the following inequality also hold
\begin{equation}
\label{eq_Density_new_Necc_1}
\begin{array}{lll}
z^{\rm T} F z
- (B+C)\rho(x) < 0,
\end{array}
\end{equation}
where $F = \begin{pmatrix}
H^{\rm T}Q+QH & BH \\
* & -1
\end{pmatrix}$.
Since $div\{\rho(x)f(x)\}<0$, then the first condition (i) in Theorem \ref{Th2} is satisfied.

Now let us check whether the sufficient stability condition (see Thorem \ref{Th2a}) is satisfied.

Given \eqref{eq_Lurie} and \eqref{eq_Density}, we calculate the following expressions
\begin{equation}
\label{eq_Density_new_Case1_1}
\begin{array}{lll}
div\{\rho(x)f(x)\}-\rho(x)div\{f(x)\} = z^{\rm T} F_0 z,
\\
div\{\rho(x)f(x)\}-\rho^2(x)div\{\rho^{-1}(x)f(x)\} 
\\
= 0.5 z^{\rm T} F_0 z.
\end{array}
\end{equation}
Using the S-procedure, if the inequalities $div\{\rho(x)f(x)\}-\rho(x)div\{f(x)\}<0$ (or $div\{\rho(x)f(x)\}-\rho^2(x)div\{\rho^{-1}(x)f(x)\}<0$) and $\varphi^2(x_2) \leq 1$ are satisfied simultaneously, then the following inequality holds
\begin{equation}
\label{eq_Density_new_Case1_2}
\begin{array}{lll}
z^{\rm T} F z < 0.
\end{array}
\end{equation}

For $F<0$, we have $div\{\rho(x)f(x)\}-\rho(x)div\{f(x)\}<0$ and $div\{\rho(x)f(x)\}-\rho^2(x)div\{\rho^{-1}(x)f(x)\}<0$. 
Therefore, the conditions (i) and (iii) of Theorem \ref{Th2a} are satisfied.


\subsection{Numerical Studies of Results in Section \ref{Density2}}

\begin{figure}[h]
\begin{minipage}[h]{0.9\linewidth}
\center{\includegraphics[width=1\linewidth]{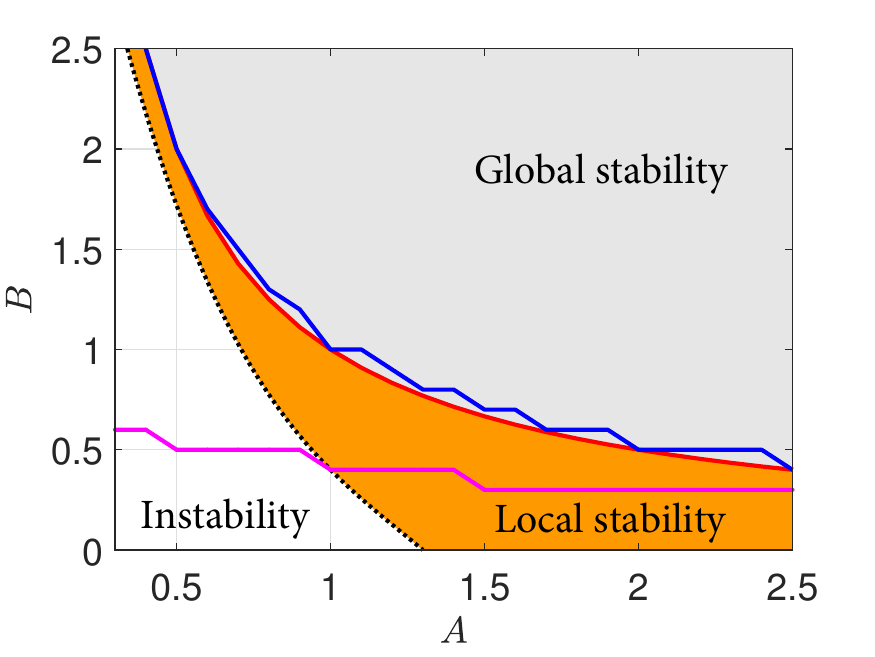}}
\end{minipage}
\vfill
\begin{minipage}[h]{0.9\linewidth}
\center{\includegraphics[width=1\linewidth]{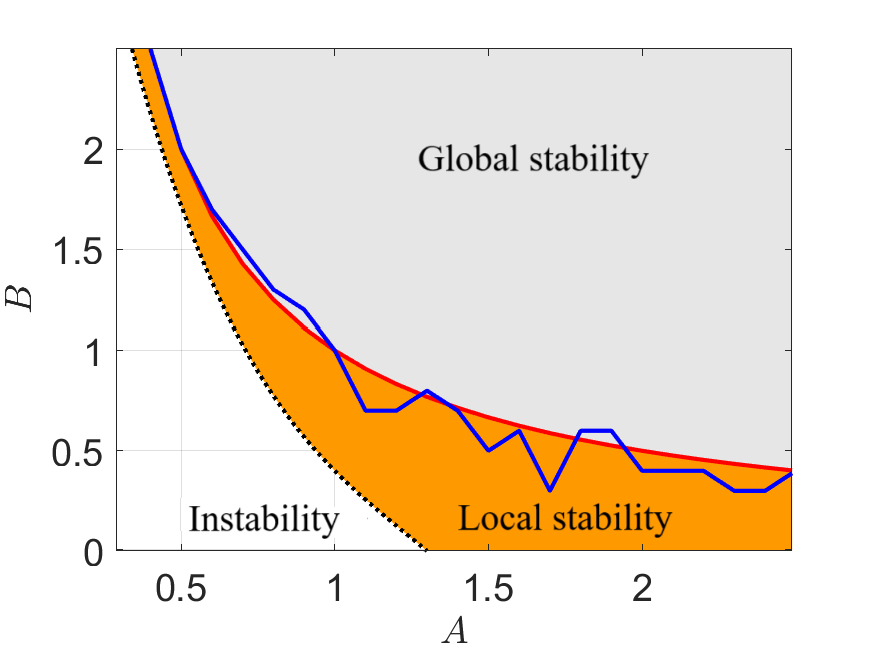}}
\end{minipage}
\caption{Deviation of the calculated global stability boundary from $AB=1$ with $C=0$ and $\alpha=0$.}
\label{Fig_LMI_A_B_1}
\end{figure}

If $C=0$ and $\alpha=0$, then the global stability region obtained by simulation practically coincides with the theoretical result $AB>1$ (see Fig.~\ref{Fig_LMI_A_B_1}, top).
Also in Fig.~\ref{Fig_LMI_A_B_1}, top, the purple line (and the region above) corresponds to the solution of the necessary condition \eqref{eq_Density_new_Necc_1}.
The CVX and Yalmip/SeDuMi software packages (solvers) are used to solve the LMI.
As the solver's computational accuracy increases, the global stability region shifts into the local stability region (see Fig. ~\ref{Fig_LMI_A_B_1}, bottom), but does not enter the instability region.

\begin{figure}[h]
\begin{minipage}[h]{0.9\linewidth}
\center{\includegraphics[width=1\linewidth]{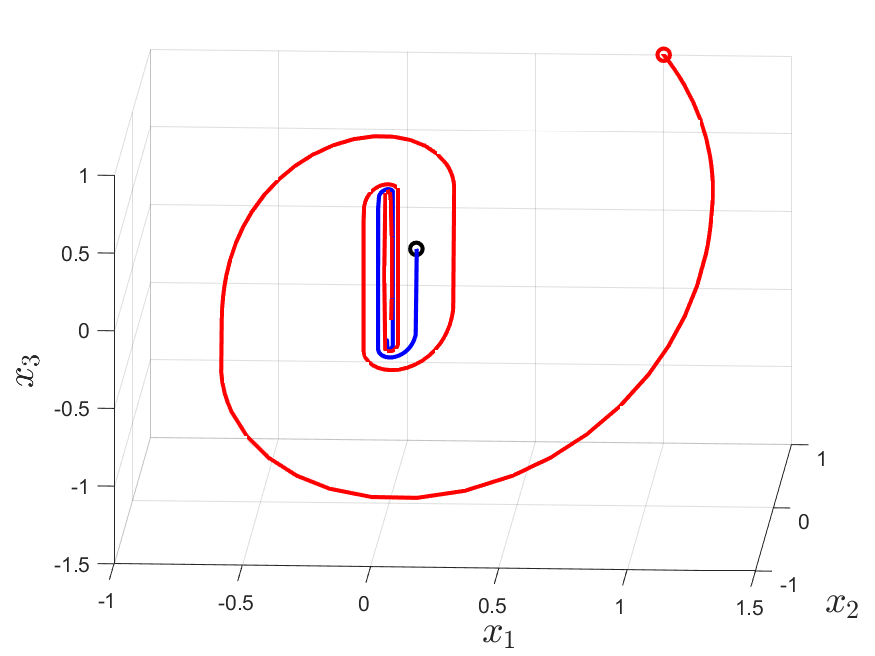}}
\end{minipage}
\vfill
\begin{minipage}[h]{0.9\linewidth}
\center{\includegraphics[width=1\linewidth]{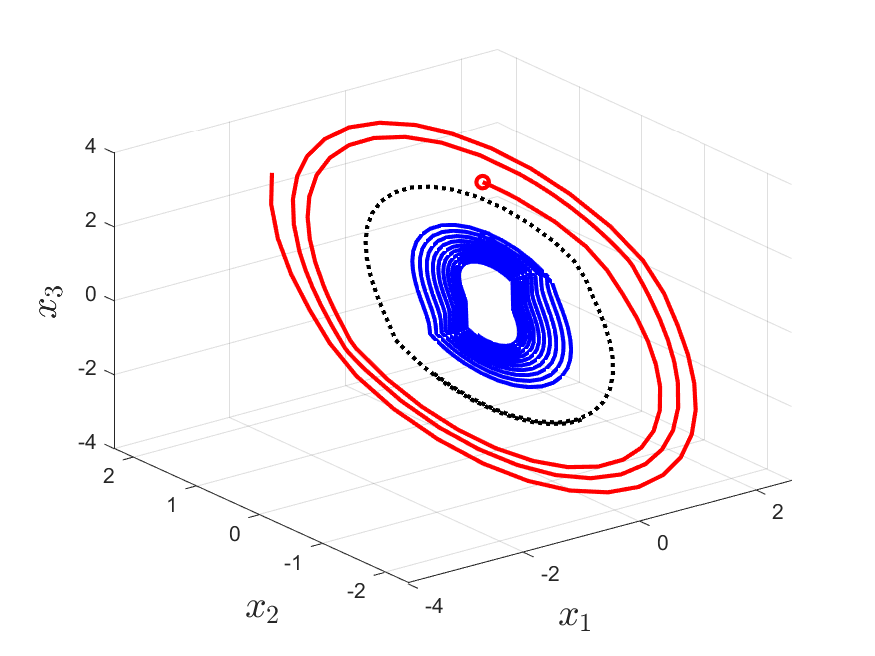}}
\end{minipage}
\caption{Phase trajectories with $B=1$ and $C=0$, as well as $A=1.4$ (top) and $A=0.75$ (bottom).}
\label{Fig_LMI_A_B_2}
\end{figure}

Fig.~\ref{Fig_LMI_A_B_2} shows phase portraits for a point taken from the global stability region ($A=1.4$, $B=1$ and $C=0$) and the region of local stability ($A=0.75$, $B=1$ and $C=0$).

\begin{figure}[h]
\center{\includegraphics[width=0.9\linewidth]{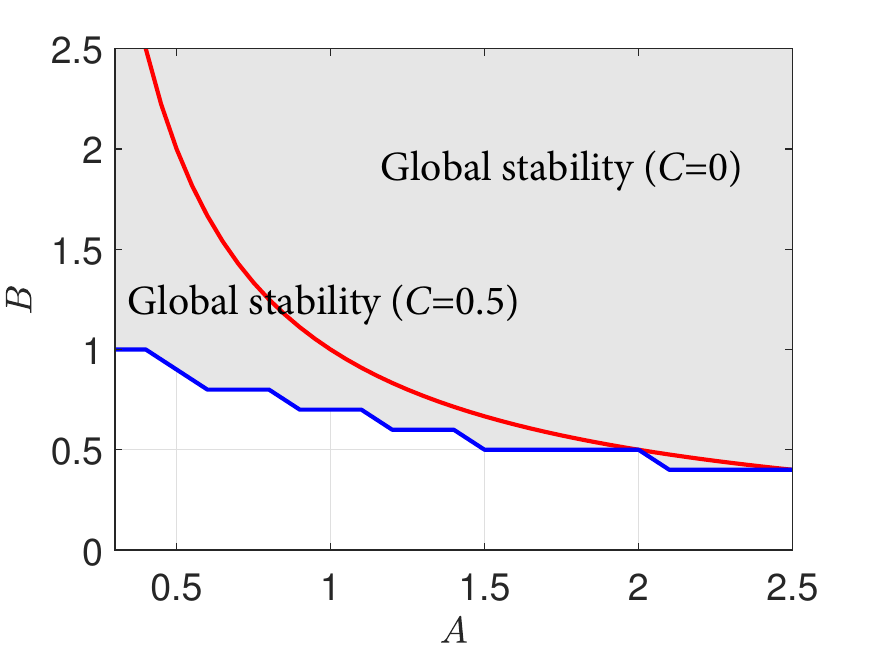}}
\caption{Deviation of the calculated global stability boundary from $AB=1$ with $C=0.5$ and $\alpha=1$.}
\label{Fig_LMI_A_B_C_1}
\end{figure}

If $C=0.5$ and $\alpha=1$, then the global stability region, obtained by simulation, increases due to the self-regulation effect in the Watt regulator (see Fig.~\ref{Fig_LMI_A_B_C_1}).

\begin{figure}[h]
\begin{minipage}[h]{0.9\linewidth}
\center{\includegraphics[width=1\linewidth]{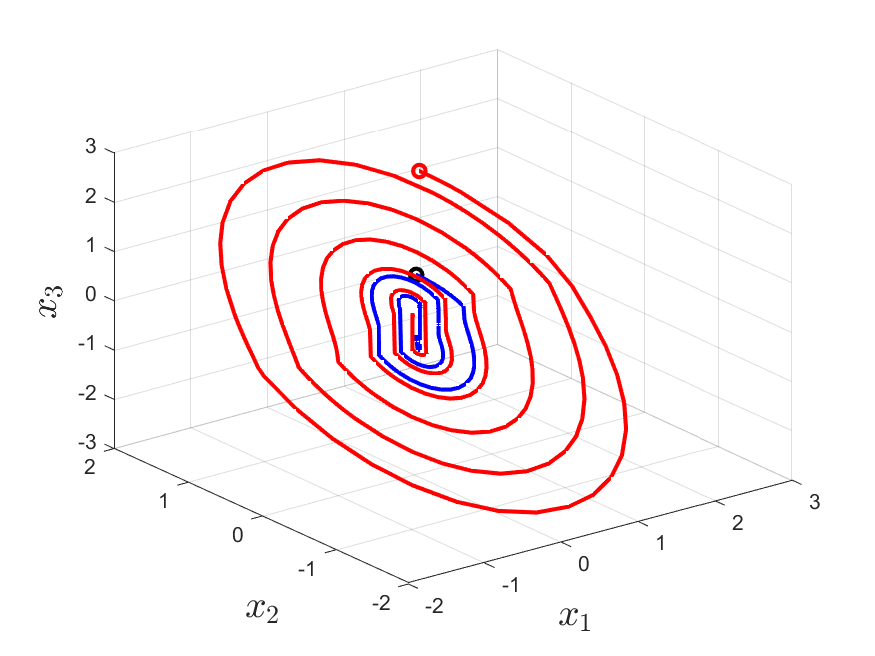}}
\end{minipage}
\hfill
\begin{minipage}[h]{0.9\linewidth}
\center{\includegraphics[width=1\linewidth]{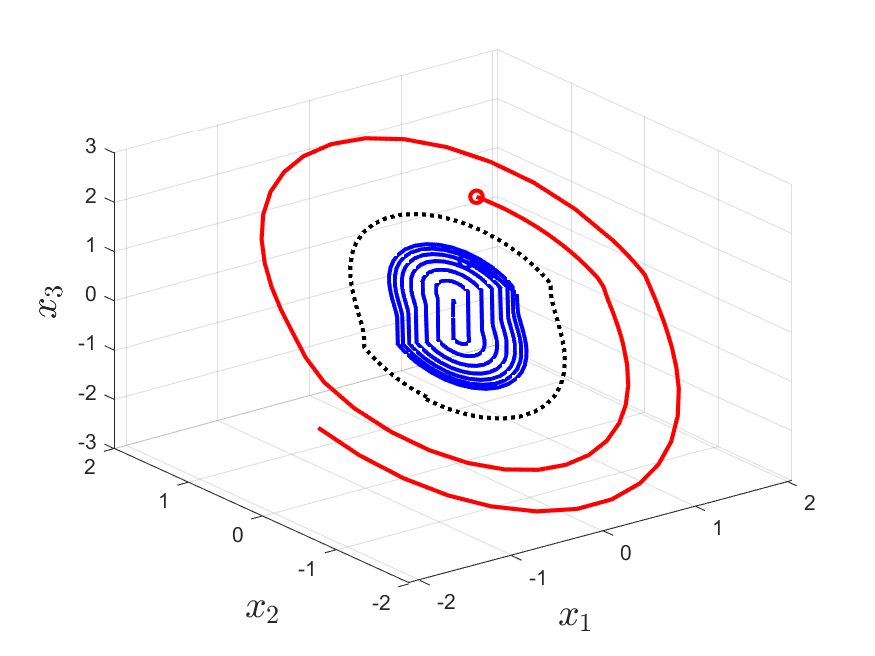}}
\end{minipage}
\caption{Phase trajectories with $B=1$ and $C=0.1$, as well as $A=0.75$ (top) and $A=0.5$ (bottom).}
\label{Fig_LMI_A_B_C_00}
\end{figure}

Fig.~\ref{Fig_LMI_A_B_C_00} shows phase portraits for a point taken from the global stability region ($A=0.75$, $B=1$, and $C=0.1$) and the local stability region ($A=0.5$, $B=1$, and $C=0.1$).
It is clear that due to the emergence of the self-regulation effect ($C > 0$), the system \eqref{eq_Lurie} is stable for the same parameters $A$ and $B$ for which it is was locally stable at $C=0$.
Moreover, the \eqref{eq_Lurie} system is locally stable for $A=0.5$, $B=1$, and $C=0.1$, whereas it is unstable for $A=0.5$, $B=1$, and $C=0$.

\begin{figure}[h]
\center{\includegraphics[width=0.9\linewidth]{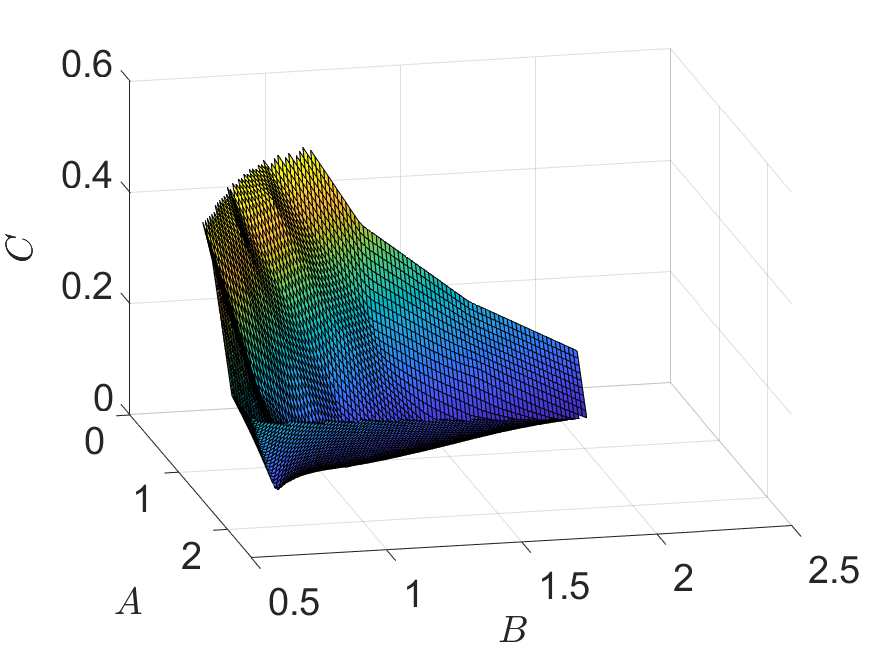}}
\caption{Global stability boundary for parameters $A$, $B$, and $C$.}
\label{Fig_Surf}
\end{figure}

Fig.~\ref{Fig_Surf} shows the global stability boundary for three parameters $A$, $B$, and $C$, calculated under the condition that $F<0$.

\section{Conclusion}
\label{Concl}

This paper applies a divergent method to the stability of dynamic systems to solve the classical Andronov-Vyshnegradsky problem, which is related to finding stability regions in systems with a Watt regulator.
This system is studied in the absence and presence of the self-regulation effect.
Precise boundaries of the hidden global stability region are obtained.
Stability criteria for a system with a Watt regulator are also proposed in the form of the solvability of the LMI.
It is shown that hidden oscillations exist in the system in the absence and presence of the self-regulation effect.
The presence of the self-regulation effect "increases" the size of the global stability region compared to the absence of this effect.





\end{document}